\documentstyle[aas2pp4]{article}
\lefthead{Balogh et al.}
\righthead{}

\begin{document}
\newcommand{\oii}{[OII]$\lambda$3727}
\newcommand{\ow}{$W_{\circ}(OII)$}
\newcommand{\mow}{$\overline{W_{\circ}}(OII)$}
\title{The Dependence of Cluster Galaxy Star Formation Rates on the Global Environment}

\author{Mike L. Balogh\altaffilmark{1}, 
David Schade\altaffilmark{2, 5}, 
Simon L. Morris\altaffilmark{2, 5},}
\author{H. K. C. Yee\altaffilmark{3, 5},
R.G. Carlberg\altaffilmark{3, 5}, 
and Erica Ellingson\altaffilmark{4, 5}}
\altaffiltext{1}{\small{Department of Physics \& Astronomy, University of Victoria, Victoria, BC, V8X 4M6, Canada. \\ email: balogh@uvastro.phys.uvic.ca}}

\altaffiltext{2}{\small{Dominion Astrophysical Observatory, National Research Council, 5071 West Saanich Road, Victoria, B.C., V8X 4M6 Canada. email: David.Schade, Simon.Morris@hia.nrc.ca}}

\altaffiltext{3}{\small{Department of Astronomy, University of Toronto, Toronto, Ontario, M5S 1A7 Canada. \\ email: hyee, carlberg@astro.utoronto.ca}}

\altaffiltext{4}{\small{CASA, University of Colorado, Boulder, Colorado 80309-0389. \\ email: e.elling@pisco.colorado.edu}}
\altaffiltext{5}{\small{Visiting Astronomer, Canada--France--Hawaii Telescope, which is operated by the National Research Council of Canada, le Centre Nationale de la Recherche Scientifique, and the University of Hawaii.}}

\begin{abstract}
A comparison of star formation properties as a function of environment is made from
the spectra of identically selected cluster and field galaxies in the CNOC 1 redshift 
survey of over 2000 galaxies in the fields of fifteen X--ray 
luminous clusters at $0.18<z<0.55$.  
The ratio of bulge luminosity to total galaxy luminosity (B/T) is computed
for galaxies in this sample, and this measure of morphology is
compared with the galaxy star formation rate as determined from the \oii\ emission line.
The mean star formation rate of cluster galaxies brighter than $M_r= -17.5 + 5 \log{h}$ 
is found to vary from 
$0.17 \pm 0.02 h^{-2}M_\odot \mbox{yr}^{-1}$ at $R_{200}$ (1.5--2 $h^{-1}$ Mpc) 
to zero in the cluster center, and is always less than the mean 
star formation rate of field galaxies, which is 
$0.39 \pm 0.01 h^{-2}M_\odot \mbox{yr}^{-1}$.  It is demonstrated that this 
significant difference is not due exclusively to the difference in morphological
type, as parameterized by the B/T value, by correcting for the B/T--radius relation.  
The distribution
of [OII] equivalent widths among cluster galaxies is skewed toward lower values relative to 
the distribution for field galaxies of comparable physical size, 
B/T and redshift,
with a statistical significance of more than 99\%.
The cluster environment affects not only the morphological mix of the galaxy population,
but also suppresses the star formation rate within those galaxies, 
relative to morphologically similar galaxies in the field.
\end{abstract}


\section{Introduction} \label{sec-intro}
In the hierarchical cold
dark matter model of structure formation, galaxy clusters form by the collapse
of rare, highly overdense regions of the universe and
continually accrete mass from the surrounding regions as they evolve.  
The cluster population may
consist of galaxies that formed long ago within the high density region, as well as those
that formed in the less dense field environment and were subsequently
accreted.  In either case, there is ample reason
to suspect that galaxies within clusters will be quite different from those
in the field.  A galaxy that forms in a high density
region of the universe may be subjected to harassment and distortion from nearby
protogalaxies (\cite{Dress}; \cite{WGJ}; \cite{G+98}), while 
one that forms in a low density environment
and is subsequently accreted may be subjected to tidal forces due to
the large gravitational potential, or interactions with other galaxies and the
intra--cluster medium (\cite{BD86}; \cite{GJ}; \cite{BV}; \cite{BH}; \cite{M+96};
 \cite{G+98}; \cite{MLK}).

It is not surprising, then, that cluster galaxies are observed to differ 
from field galaxies in their morphologies, colors and star formation rates 
(e.g., \cite{Dress}; \cite{WGJ}; \cite{A2390}; \cite{D+97}; \cite{B+97}; 
\cite{H+97}; \cite{F+98}; \cite{KK}; \cite{1621}).  
Although it has been shown that the mean star formation rate (SFR) in cluster
galaxies is always less than in field galaxies (\cite{B+97}), it is important to 
determine if this remains true when the different morphological 
composition of the two populations is accounted for, since SFR
is known to be strongly correlated with morphological type (e.g., \cite{K92}). 
This will distinguish between two simple scenarios: 1) the cluster environment
favors galaxies of a certain morphological type, in terms of their size and
relative bulge and disk components, but the SFR of these galaxies
is statistically equivalent to that of similar galaxies in the field; or, 2)
the SFR of a galaxy depends on its environment as well as its morphology.  The
results of this work will provide strong support for the second hypothesis.

The plan of this paper is as follows.  In \S~\ref{sec-sample} the data sample is 
defined, selection effects are considered, and the measurement techniques are described.  
In \S~\ref{sec-res} the emission line properties and morphologies of cluster galaxies are 
compared with the field sample.  
Implications are discussed in \S~\ref{sec-discuss}, and 
the conclusions are summarized in \S~\ref{sec-conc}.
A cosmology of $q_{\circ}=0.1$ is assumed for distance dependent 
calculations, which are given in terms of $h=H_{\circ}/100$ throughout. 

\section{Sample Selection and Measurements}\label{sec-sample}
The CNOC 1 cluster sample\footnote{The measured parameters discussed here,
as well as others and the raw data itself, will soon be available from the CADC data
archive.} consists of fifteen\footnote{Omitting 
cluster E0906+11, for which a velocity dispersion could not be computed (\cite{CNOC1}).}
 X--ray luminous clusters, observed with MOS at CFHT, in the redshift
range $0.18<z<0.55$.  Redshifts were obtained for about 2500 galaxies, and observations
 extend as far out as 1--2 $R_{200}$ in projected distance for most clusters,
where $R_{200}$ is the radius at which the mean interior mass density is equal to
200 times the critical density, and within which it is expected that the galaxies are
in virial equilibrium (\cite{GG}; \cite{CER}).  
The observational strategy and details of the survey 
are detailed in Yee, Ellingson \& Carlberg (1996).
Cluster members are considered to be those galaxies with velocity differences from the
brightest cluster galaxy (BCG\footnote{Except for cluster MS 0451.5+0250, for which no
redshift is available for the BCG.  The velocities are measured relative to the
mean for this cluster.}) less than 3$\sigma(r)$, where $\sigma(r)$ is the cluster 
velocity dispersion as a function of projected radius $r$, as determined from the mass
models of Carlberg, Yee \& Ellingson (1997).  Field galaxies are selected to be those
with velocities greater than 6$\sigma(r)$.  The cluster-centric distance parameter $R$ is the 
projected distance from the cluster
BCG, and will be normalized to $R_{200}$, since there is a small range in 
the mass and linear size of the clusters in this sample.

The equivalent width of the \oii\ emission line, \ow, was automatically computed by summing the
observed flux above the continuum in pixels between $3713<\lambda< 3741$ \AA.  The 
continuum level was estimated by fitting a straight line to the flux between $3653 <\lambda 
< 3713$ \AA\ and $3741<\lambda < 3801 $ \AA\ using weighted linear regression, with weights 
from the Poisson noise vector generated by optimally extracting the spectra with 
IRAF\footnote{IRAF is distributed by the National Optical Astronomy Observatories which is 
operated by AURA 
Inc. under contract with NSF.}.  The error in \ow\ is computed from equation A8 in 
Bohlin et al. (1983) and an average \ow, weighted by this error, is adopted for multiply 
observed galaxies in the sample.  The mean and median error in \ow\ is 5 \AA\ and 3 \AA, 
respectively, for the full sample.  These error estimates were found to be reasonably
representative of the reproducibility of multiple \ow\ measurements, as described in 
Balogh et al. (1997).  

Morphological parameters for the Gunn r band MOS images were measured by fitting two
dimensional models of exponential disk and $R^{1/4}$ law profiles to the symmetrized components
of the light distribution, as described in Schade et al. (1996a, 1996b).  The images are
symmetrized to minimize the effects of nearby companions and asymmetric structure, and a
$\chi^2$ minimization procedure is applied to the models, convolved with the image point
spread function, to obtain best fit values of the galaxy size, surface brightness and fractional
bulge luminosity (bulge--to--total, or B/T ratio).  
Simulations show that the B/T measurements
are reliable within about 20\% for images of this quality (\cite{Schade1}).  

The data are weighted by two factors: a 
magnitude weight $W_m$ which compensates for the fact that it is more difficult to 
obtain redshifts
for faint galaxies, and $W_{ring}$, which corrects for non--uniform sampling as a function
of distance from the cluster center (\cite{YEC}).  The latter weight is only important when
global cluster properties are considered such that the properties of galaxies at large
radii are averaged together with those at small radii.  Good fits to the 
light profiles were obtained for 1143 (712 cluster, 373 field, 58 near--field) 
of the 1515 galaxies with  $W_m \le 5$ and errors in \ow\ of less than 10 \AA, and 
these comprise the selected subsample\footnote {The BCGs
are also omitted from the sample, as they are clearly non--typical cluster members.
  Only eight of the fourteen BCGs with redshifts are well fit by the simple
two component model light profile, due to significant crowding and superposition of 
galaxies near the center of the cluster, and  
eight show strong emission lines, indicating very strong SFRs of 
between 1.8 and 21 $h^{-2} M_\odot \mbox{yr}^{-1}$.}.   
The emission line properties and relative abundance 
of the galaxy population with poorly fit luminosity profiles are not
 significantly correlated with radial distance or cluster membership, and thus the 
exclusion of these galaxies from the sample is unlikely to bias the results.  
The selected sample includes galaxies with absolute Gunn r 
magnitudes less than about $-17.5 + 5 \log{h}$, 
and is complete to $M_r=-18.5+5 \log{h}$.  The absolute
magnitude distribution of the cluster sample is not significantly different from that of 
field.  Furthermore, the absolute magnitude distribution is similar for the low
and high redshift galaxies, a result of the longer exposure times in the high
redshift cluster images (\cite{YEC}).  Although a proper treatment of the redshift evolution 
is not the present focus, it is noted that there is no significant difference in the results 
of this investigation 
between the low and high redshift clusters in the sample. 

The sample is divided into three classes (hereafter referred to as B/T classes)
based on the measured B/T value. 
The bulge dominated class (B) consists of those galaxies with $B/T>0.7$, the disk
dominated class (D) those with $B/T<0.4$, and intermediate galaxies are classified
``Int''.  These classes should not be confused with or forced to conform to more familiar
Hubble types, which depend partly on star formation properties and 
the presence of spiral structure.
The B/T ratio and the sizes of the disk and bulge components are more stable properties
that reflect the true ``morphology'' of the galaxy and are less dependent on its star
formation properties.  However, the measured B/T may somewhat underestimate the ``intrinsic''
(i.e. representative of the mass distribution) value in star forming galaxies, 
as star formation takes place preferentially in the disk.

\section{Results}\label{sec-res}
\begin{figure}
\plotone{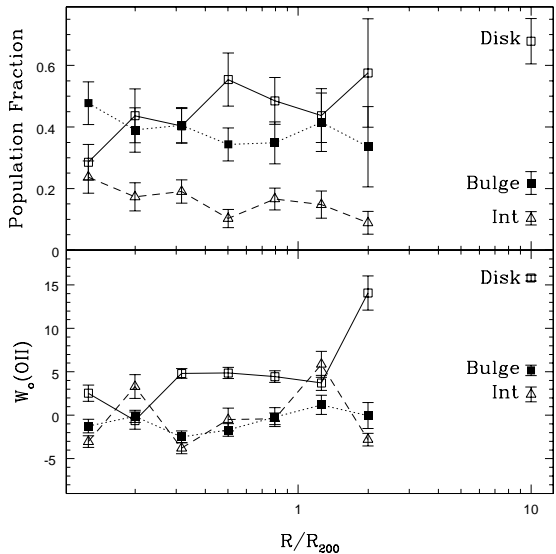}
\figcaption{The top panel shows the fraction of galaxies in each
B/T class as a function of cluster-centric radius.  Open squares,
connected by the solid line, represent the D class ($B/T<0.4$).  The solid 
squares (dotted line) represent the B class ($B/T>0.7$), and the 
triangles (dashed line) the Int class.  The field values are plotted at $R/R_{200}=10$ for
display purposes only.  Shown in the bottom panel is the
weighted observed \mow\ for each class of galaxy as a function of radius, with the field
values again plotted at $R/R_{200}=10$.
\label{fig-oiirad_full}}
\end{figure}

The B/T--radius relation for the cluster sample
is shown in the top panel of Figure \ref{fig-oiirad_full}.   The fraction of field galaxies in
each B/T class is shown at $R/R_{200}=10$, strictly for display purposes.  The radial bin
sizes are equal in logarithmic intervals, except for the innermost bin, which represents
all galaxies at $R<0.16R_{200}$.  
The fraction
of disk dominated galaxies, $f_D(r)$, decreases fairly steadily toward the center of the cluster,
as expected, from about 70\% in the field to 30\% in the cluster center.  The radial gradients
of the B and Int populations ($f_B(r)$, $f_{Int}(r)$) are nearly identical, 
and their proportions increase by about
15\% between the outer and central cluster regions.   
The bottom panel of Figure \ref{fig-oiirad_full} shows the mean \ow\ for galaxies in each 
B/T class as a function of radius.  The field value, again shown at $R/R_{200}=10$, for 
galaxies of class B, Int, and D is \mow $_B=5.1 \pm 0.6$ \AA, 
\mow $_{Int}=2.1 \pm 0.9$ \AA, and \mow $_D=15.8 \pm 0.3$ \AA\
respectively.  The non--zero \mow\ for the B class is significant
at the $3\sigma$ level, and many B galaxies have quite strong emission lines 
(10\% have \ow\ $>14.0$ \AA).  Thus, signs of significant star formation are found
in galaxies with little or no disk component, though this may partly reflect the
uncertainty in B/T.  Within the cluster the \mow\ for B and Int class
galaxies is consistent with zero, and there is little variation with radius.  
The \mow\ of D galaxies is only consistent with the
field at 2$R_{200}$; at smaller radii it is always less than the field value by at least 10 \AA.
The value of \mow\ depends not only on the B/T parameter, but also on the
environment, in the sense that it is lower for galaxies in clusters than for galaxies with the
same B/T ratio in the field.

\begin{figure}
\plotone{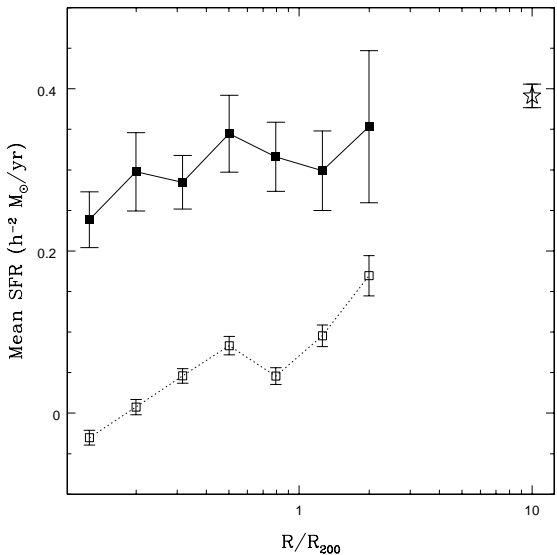}
\figcaption{The weighted observed $\overline{SFR}$ (dotted line, open symbols) 
as a function of radius.  
The field value is represented as the large star
at $R/R_{200}=10$.  The solid points (connected by the solid line) represent the
$\overline{SFR}$ that would be observed at each radius if the $\overline{SFR}$ 
of galaxies in the B, Int and
D classes was equal to its corresponding value in the field, $\overline{SFR}_B=0.14 \pm 0.02$,
$\overline{SFR}_{Int}=0.10 \pm 0.02$ and $\overline{SFR}_{D}=0.52 \pm 0.02$, in units of $h^{-2} M_\odot \mbox{yr}^{-1}$.
(see \S~\ref{sec-res} for details).  
\label{fig-sfr}}
\end{figure}

The actual SFR of a galaxy is directly related to the luminosity
of the [OII] emission line (\cite{G+89}; \cite{K92}; \cite{BP}), 
although the constant of proportionality is somewhat uncertain.  The relation proposed by 
Barbaro \& Poggianti (1997) 
will be used  here, but since the present concern is the relative
SFR of cluster and field galaxies, the constant of proportionality
is unimportant.  The luminosity of the [OII] line is calculated from the equivalent width
and the galaxy's rest frame B magnitude following Kennicutt's (1992) relation, 
with the suggested extinction at $H\alpha$ of 1 magnitude.
The mean SFR for field galaxies in each B/T
class is $\overline{SFR}_B=0.14 \pm 0.02$,
$\overline{SFR}_{Int}=0.10 \pm 0.02$ and $\overline{SFR}_{D}=0.52 \pm 0.02$, 
in units of $h^{-2} M_\odot \mbox{yr}^{-1}$.  If cluster galaxies of a given B/T type had
identical $\overline{SFR}s$ to corresponding field galaxies, then the mean cluster
SFR at radius r could be determined from the relation
$\overline{SFR}(r)=f_B(r)\times\overline{SFR}_B+f_{Int}(r)\times\overline{SFR}_{Int}+
f_D(r)\times\overline{SFR}_D$.  This ``predicted'' relation is shown as the solid
line in Figure \ref{fig-sfr}.  The error bars displayed include both the error in the 
$\overline{SFR}$ values of each B/T class, and the uncertainty in the population fractions 
in each radial bin.  The observed $\overline{SFR}$ is shown as the 
dotted line; it varies by 
$0.2 h^{-2} M_\odot \mbox{yr}^{-1}$ over the observed radial range, 
but for $R<R_{200}$ it is always less than the field value (corrected for the B/T--radius 
relation) by more than $3\sigma$.
At $2R_{200}$ the cluster galaxies still have lower star formation by more than a
factor of two, although the difference is only significant at the 1.9 $\sigma$ level.
This suggests that large changes in SFR may occur well outside
the virial radius. 

\section{Discussion} \label{sec-discuss}
\begin{figure}
\plotone{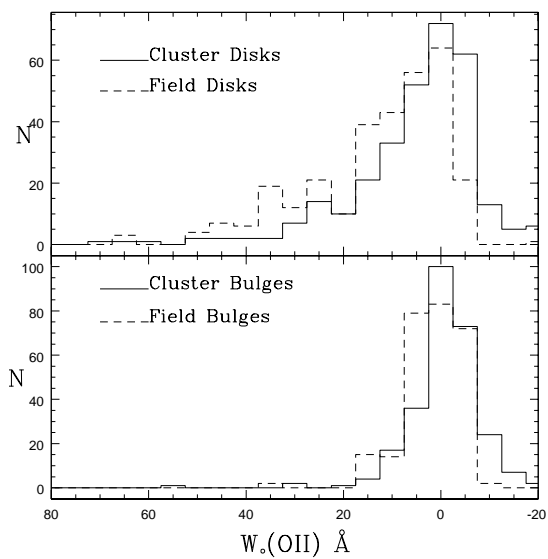}
\figcaption{The top panel shows the \ow\ distributions for D galaxies in the
cluster (solid line) and a morphologically matched field sample (dashed line, 
see \S~\ref{sec-discuss} for
details).  Similar distributions are shown for B galaxies in the bottom
panel.  For both B and D galaxies, the difference in \ow\ distributions 
between morphologically analagous cluster and field galaxies 
is significant with more than 99\% confidence as determined by a standard K--S test.
\label{fig-dist}}
\end{figure}

It was shown in Balogh et al. (1997) that, on average, cluster galaxies have less star formation
than field galaxies, and that there is no evidence for ongoing star formation in 
excess of the field at any radius.  Figures \ref{fig-oiirad_full} and \ref{fig-sfr} show  
that this result cannot be accounted for by assuming a universal dependence of SFR on B/T.  
The B/T measure, however, is not sufficient to characterize a galaxy's physical structure,
as there is a large range of physical bulge and disk sizes for a given B/T.  To account for
a dependence of \ow\ on galaxy size, cluster galaxies are compared with an
analogous field sample in the following manner.  For every class D cluster galaxy, 
a corresponding field galaxy is found which has a similar redshift, B/T, and disk scale length.
Only those galaxies for which a match exists are considered.  Thus,  a cluster and matched
field sample are chosen such that the
only significant difference in the selection of the two samples 
is the global environment of the galaxies.  The \ow\ distributions of the
cluster and matching field disk sample are shown in the top panel of Figure \ref{fig-dist}.

A similar procedure is followed for the B class galaxies,
where the field galaxies are chosen to match the cluster
sample in redshift, B/T and physical bulge size; these distributions are shown in the bottom
panel.  For both B and D objects, the hypothesis that the cluster and field distributions 
of \ow\ are drawn from the same population is rejected with 
more than 99\% significance by a standard Kolmogorov--Smirnov test.  
For the B galaxies, the difference is due to an excess of field galaxies with weak emission
lines (\ow$<10$ \AA) relative to the cluster; few galaxies are seen in either the cluster or 
field with
stronger lines.  The D galaxies, on the other hand, show a significant excess of emission lines
of all strengths in the field, relative to the cluster.  In the mean,
galaxies in clusters have lower SFRs
than galaxies in the field, independent of their B/T ratio or physical size.

Figure \ref{fig-sfr} indicates that the mean star
formation rate in cluster galaxies is still lower than that in the field around 2 $R_{200}$, 
which is approximately the virial radius in an $\Omega=0.2$ universe.  If cluster 
galaxies have undergone strong bursts of star formation in the past 1--2 Gyr,
as suggested by several authors (e.g., \cite{CS87}; \cite{MW93}; \cite{Barger}; 
\cite{C+96}; \cite{K+97}; \cite{C+98}), then the elusive population of starbursting
galaxies may be present beyond this radius.  Ram pressure or tidal 
stripping may still be viable mechanisms for reducing the star formation in these galaxies
without a burst if they have already passed through the cluster center at least once,
a scenario for which there is some support (\cite{G+98}; \cite{RdS}).
  
Alternatively, the star formation properties of galaxies may have been altered  
before they became bound to the cluster.  In particular,
Hashimoto et al. (1997) and Couch et al. (1997) suggest that galaxy--galaxy interactions
and mergers are responsible for inducing starbursts in lower density environments.  Such
interactions would be favored in galaxy groups, which have relatively low
velocity dispersions.  In the hierarchical model of structure formation, groups will 
merge to form clusters, and it may be that the reduced star formation observed among some 
cluster galaxies today is the result of their previous existence in a group environment. 
In this case, the population of strongly star forming galaxies may be found in such
groups, and not within clusters at all.

\section{Summary}\label{sec-conc}
The result of Balogh et al. (1997), that current star formation among cluster galaxies
is suppressed relative to an identically selected field sample, is not simply a reflection
of a different morphological composition of the two populations.  Although the fraction 
of galaxies with a significant
disk component decreases from about 70\% in the field to 30\% in the cluster center,
the $\overline{SFR}$ of cluster galaxies is much less
than can be explained by this correlation alone, significant at more than the $3\sigma$ level.  
Cluster galaxies have less star formation than field 
galaxies of similar physical size, fractional bulge luminosity and redshift.  Even at
the virial radius, around 2 $R_{200}$, the amount of star formation in cluster
galaxies is less than that in the field, which suggests either that the global 
environment is affecting galaxy star formation at or beyond this radius (or within
a group environment), or that a
significant number of cluster galaxies near the virial radius have already passed
through or near the cluster interior.   
  
\acknowledgments 
MLB is supported by the Natural Sciences and Engineering Research Council of Canada.


\clearpage

\begin{thebibliography}{}
\bibitem[Abraham et al. 1996]{A2390} Abraham, R. G. et al. 1996 \apj, 471, 694
\bibitem[Balogh et al. 1997]{B+97}Balogh, M. L., Morris, S. L., Yee, H. K. C., Carlberg, R. G. \& Ellingson, E. 1997, \apj, 488, L75
\bibitem[Barbaro \& Poggianti 1997]{BP}Barbaro, G. \& Poggianti, B. M. 1997, \aap, 324, 490
\bibitem[Barger et al. 1996]{Barger}Barger, A. J., Arag\'{o}n-Salamanca, A., Ellis, R. S., Couch, W. J., Smail, I., \& Sharples, R. M. 1996, \mnras, 279, 1
\bibitem[Barnes \& Hernquist 1991]{BH}Barnes, J. E., \& Hernquist, L. E. 1991, \apj, 370, L65
\bibitem[Bohlin et al. 1983]{Bohlin} Bohlin, R. C., Hill, J. K., Jenkins, E. B., Savage, B. D., Snow,T. P. Jr., Spitzer, L. Jr., \& York, D. G. 1983, \apjs, 51, 277
\bibitem[Bothun \& Dressler 1986]{BD86}Bothun, G. D., \& Dressler, A. 1986, \apj, 301, 57
\bibitem[Byrd \& Valtonen 1990]{BV}Byrd, G., \& Valtonen, M. 1990, \apj, 350, 89
\bibitem[Caldwell et al. 1996]{C+96}Caldwell, N., Rose, J. A., Franx, M. \& Leonardi, A. J. 1996, \aj, 111, 78
\bibitem[Carlberg, Yee \& Ellingson 1997]{CYE}Carlberg, R. G., Yee, H. K. C. \&  Ellingson, E. 1997, \apj, 478, 462
\bibitem[Carlberg et al. 1996]{CNOC1} Carlberg, R. G., Yee, H. K. C., Ellingson, E., Abraham, R., Gravel, P., Morris, S., \& Pritchet, C. J. 1996, \apj, 462, 32
\bibitem[Couch \& Sharples 1987]{CS87}Couch, W. J., \& Sharples, R. M. 1987, \mnras, 229, 423
\bibitem[Couch et al. 1998]{C+98}Couch, W. J., Barger, A. J., Smail, I., Ellis, R. S. \& Sharples, R. M. 1998, \apj, 497, 188
\bibitem[Crone, Evrard \& Richstone 1994]{CER}Crone, M. M., Evrard, A. E., Richstone, D. O. 1994, \apj, 434, 402
\bibitem[Dressler 1980]{Dress}Dressler, A. 1980, \apj, 236, 351
\bibitem[Dressler et al. 1997]{D+97}Dressler, A., Oemler, A. Jr., Couch, W. J., Smail, I., Ellis, R. S., Barger, A., Butcher, H., Poggianti, B. M. \& Sharples, R. M. 1997, \apj, 490, 577
\bibitem[Fisher et al. 1998]{F+98}Fisher, D., Fabricant, D., Franx, M. \& van Dokkum, P. 1998, \apj, 498, 195
\bibitem[Gallagher et al. 1989]{G+89}Gallagher, J. S., Bushouse, H. \& Hunter, D. A. 1989, \aj, 97, 700
\bibitem[Gavazzi \& Jaffe 1987]{GJ}Gavazzi, G., \& Jaffe, W. 1987, \apj, 310, 53
\bibitem[Ghigna et al. 1998]{G+98}Ghigna, S., Moore, B., Governato, G., Lake, G., Quinn, T. \& Stadel, J. 1998, astro-ph/9801192, to appear in \mnras.
\bibitem[Gott \& Gunn 1972]{GG} Gott, J. R., \& Gunn, J. 1972, \apj, 176, 1
\bibitem[Hashimoto et al. 1997]{H+97}Hashimoto, Y., Oemler, A. Jr., Lin, H. \& Tucker, D. L. 1997, astro-ph/9712319
\bibitem[Kennicutt 1992]{K92}Kennicutt, R. C., Jr. 1992, \apj, 388, 310
\bibitem[Koo et al. 1997]{K+97}Koo, D. C., Guzm\'{a}n, R., Gallego, J. \& Wirth, G D. 1997, \apj, 478, 49
\bibitem[Koopmann \& Kenney 1998]{KK}Koopmann, R. A. \& Kenney, J. D. P. 1998, \apj, 497, 75
\bibitem[Moore et al. 1996]{M+96}Moore, B., Katz, N., Lake, G., Dressler, A. \& Oemler, A. 1996,  \nat, 379, 613
\bibitem[Moore, Lake \& Katz 1998]{MLK}Moore, B., Lake, G. \& Katz, N. 1998, \apj, 495, 139
\bibitem[Morris et al. 1998]{1621}Morris, S. L., Hutchings, J. B., Carlberg, R. G., Yee, H. K. C., Ellingson, E., Balogh, M. L., Abraham, R. G. \& Smecker--Hane, T. A. 1998, astro-ph/9805216, to appear in \apj.
\bibitem[Moss \& Whittle 1993]{MW93}Moss, C., \& Whittle, M. 1993, \apj, 407, L17
\bibitem[Ram\'{\i}rez \& de Souza]{RdS}Ram\'{\i}rez, A. C. \& de Souza R. E. 1998, \apj, 496, 693
\bibitem[Schade et al. 1996a]{Schade1}Schade, D., Lilly, S. J., LeF\`{e}vre, O., Hammer, F. \& Crampton, D. 1996a, \apj, 464, 79
\bibitem[Schade et al. 1996b]{Schade2}Schade, D., Carlberg, R. G., Yee, H. K. C., L\'{o}pez--Cruz, O., \& Ellingson, E. 1996b, \apj, 464, L63
\bibitem[Whitmore, Gilmore \& Jones 1993]{WGJ} Whitmore, B. C., Gilmore, D. M., \& Jones, C. 1993, \apj, 407, 489
\bibitem[Yee et al. 1996]{YEC} Yee, H. K. C., Ellingson, E., \& Carlberg, R. G. 1996, \apjs, 102, 629 
\end{thebibliography}
\end{document}